\documentclass[aps,preprint,nofootinbib,superscriptaddress,prc]{revtex4}
\usepackage{amssymb}
\usepackage{CJK}
\usepackage{indentfirst}
\usepackage{amsmath}
\usepackage{xcolor}

\usepackage{epsfig}
\usepackage[bookmarksnumbered,bookmarksopen,colorlinks,citecolor=blue,linkcolor=blue] {hyperref}
\begin{document}

\date{\today}

\title{Analytical penetration probability including the centrifugal potential: An improved Buck–Merchant–Perez model for 
$\alpha$-decay half-lives}

\author{Minghui Hu}
\affiliation{Department of Physics, Guangxi Normal University, Guilin 541004, People's Republic of China }

\author{Pengfei Ma}
\affiliation{Department of Physics, Guangxi Normal University, Guilin 541004, People's Republic of China }

\author{Kai Ren}
\affiliation{Department of Physics, Guangxi Normal University, Guilin 541004, People's Republic of China }

\author{Junlong Tian}
 \email{tianjl@gxnu.edu.cn}
 \affiliation{Department of Physics, Guangxi Normal University, Guilin 541004, People's Republic of China }
\affiliation{Guangxi Key
Laboratory of Nuclear Physics and Technology,  Guilin 541004, People's Republic of China}

\author{Cheng Li}
 \email{licheng@gxnu.edu.cn}
\affiliation{Department of Physics, Guangxi Normal University, Guilin 541004, People's Republic of China }
\affiliation{Guangxi Key
Laboratory of Nuclear Physics and Technology,  Guilin 541004, People's Republic of China}

\begin{abstract}
We derive a closed-form, non-perturbative WKB penetration formula for $\alpha$-decay that explicitly 
incorporates the centrifugal potential within the Buck–Merchant–Perez (BMP) cluster model. 
The centrifugal term is shown to enhance the hindrance by effectively enlarging the 
barrier width: it pushes the outer turning point outward and, via the 
Bohr--Sommerfeld quantization condition, shifts the inner turning point inward. 
Building on this analytical result, we further develop an improved BMP model in which the nuclear potential depth is expressed as a unified four-parameter formula that simultaneously encodes shell corrections, odd-even pairing effects, and orbital-angular-momentum dependence. For 534 ground-state-to-ground-state $\alpha$ decays spanning $60 \le Z \le 118$, the root-mean-square deviation of $\log_{10} T_{1/2}$ is 
reduced to 0.267, representing a 57\% improvement over the original constant-depth BMP model (0.615), with 
robust performance for both favored (0.188) and unfavored (0.398) transitions. The framework is further applied to predict the half-lives of hitherto-unmeasured nuclei in the region $Z = 117$--$120$, providing quantitative benchmarks for future experimental investigations.
\end{abstract}


\maketitle

\section{\label{sec:level1}Introduction}
$\alpha$ decay, discovered by Rutherford in 1899~\cite{Rutherford1899} and 
interpreted as quantum tunneling through the Coulomb barrier by 
Gamow~\cite{Gamow1928} (and independently by Gurney and 
Condon~\cite{Gurney1928}) in 1928, remains a paradigmatic problem at the 
interface of quantum tunneling and nuclear structure. 
Although a wide variety of theoretical models have since been developed~\cite{Xuc15,Xuc15a,CPPM,GLDM,VSS,UDL,Royer00,Royer10,NRDX,UNIV,Wangyz15}, 
Gamow’s conceptual foundation remains central: the $\alpha$ particle is assumed to be preformed inside the parent nucleus 
and to escape by penetrating the potential barrier formed by the nuclear and 
Coulomb interactions. As the most direct mathematical realization of this 
picture, the WKB approximation continues to be one of the most intuitive 
and widely used tools for evaluating $\alpha$-decay penetration probabilities. In this work, we combine an exact analytical treatment of the centrifugal barrier with an improved parametrization of the nuclear potential depth within the Buck–Merchant–Perez (BMP) framework, aiming at a unified and accurate description of 
$\alpha$-decay half-lives across the nuclear chart.

\begin{figure}
\includegraphics[angle=-0,width=1.05\textwidth]{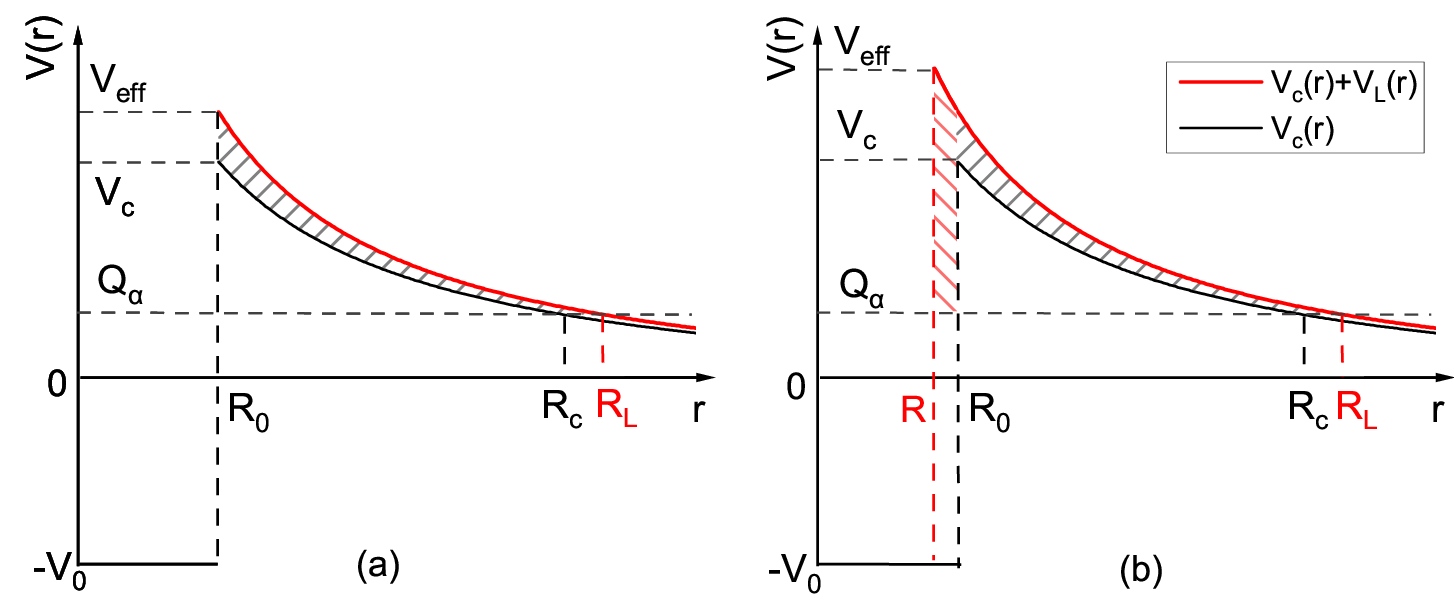}
\caption{(Color online) Schematic comparison of the $\alpha$-decay potential 
with and without the centrifugal contribution. The black and red curves 
denote the pure Coulomb barrier  $V_{\mathrm{c}}(r)$ and the effective barrier 
$V_{\mathrm{c}}(r) + V_{\mathrm{L}}(r)$, corresponding to favored ($L = 0$) and unfavored 
($L \neq 0$) decays, respectively. Including the centrifugal term raises the 
barrier height from $V_{\mathrm{c}}$ to $V_{\mathrm{eff}}$ and shifts the outer turning point 
outward from $R_{\mathrm{c}}$ to 
$R_{\mathrm{L}}$. \textbf{(a)} Perturbative treatment~\cite{Gamow1928,Zhang2011,RojasGamboa2024}, 
in which the inner turning point is fixed at its Coulomb value $R_{0}$, while 
the outer turning point is approximated by the Coulomb value 
$R_{\mathrm{c}}$. The hatched region represents the 
additional tunneling area associated with the centrifugal hindrance. \textbf{(b)} Non-perturbative treatment of the present work, where the inner turning point $R$ is determined self-consistently from the Bohr--Sommerfeld quantization condition and moves inward to $R < R_{0}$. The extra red-hatched region on the 
left indicates the additional tunneling contribution missed by the 
perturbative approach.} 
\label{fig1}
\end{figure}

As illustrated schematically in Fig.~\ref{fig1}(a), the $\alpha$ particle is 
bound in an internal nuclear well for $r < R_0$ and encounters a repulsive 
Coulomb barrier outside the nuclear surface. For favored $\alpha$ transitions, 
the emitted $\alpha$ particle typically carries zero orbital angular momentum, 
so the external barrier is governed solely by the Coulomb interaction 
$V_c(r)$ (black curve). In unfavored transitions, the spin--parity difference between the parent and daughter states 
requires a nonzero angular momentum $L$. The associated centrifugal potential, 
$V_L(r) = \hbar^{2} L(L+1)/(2\mu r^{2})$, raises the effective barrier $V_c(r) + V_L(r)$ and pushes the outer turning point further outward (red curve), thereby enlarging the WKB action and strongly suppressing the penetration probability.
The centrifugal barrier in $\alpha$ decay has traditionally been treated perturbatively, an approach first introduced by Gamow~\cite{Gamow1931} and subsequently refined by Zhang, Xu, and Ren~\cite{Zhang2011} and Rojas-Gamboa, Kelkar, and Caballero~\cite{RojasGamboa2024}. These approaches rely on the small parameter $\sigma=V_L(R_0)/V_c(R_0)\ll1$, so that the centrifugal hindrance appears as an additive contribution separable from the Coulomb tunneling factor. Notably, these derivations fix both turning points at their Coulomb values, so that only the barrier height—not its width—is modified.

In this work, we remove the small-$\sigma$ assumption and derive a 
\emph{closed-form, non-perturbative expression} valid for arbitrary $L$, in which the tunneling action for 
$V_c(r) + V_L(r)$ is evaluated exactly. The physical origin 
of the breakdown of the perturbative treatment is illustrated in 
Fig.~\ref{fig1}(b): with increasing $L$, the centrifugal potential modifies 
the barrier in two correlated ways. \textbf{(i)} The inner turning point $R$
shifts inward, as the Bohr--Sommerfeld quantization condition yields a 
nonlinear $L$-dependence that departs from Gamow’s perturbative estimate
$R_{\text{eff}} \approx R_0[1 - 0.002\,L(L+1)]$ ~\cite{Gamow1931}. \textbf{(ii)} The outer turning point shifts 
outward through the exact solution of $V(R_{\mathrm{L}}) = Q$, with the excess growing nonlinearly with $L$. The resulting 
enlargement of the barrier width produces a nonlinear amplification of 
the exponential WKB integral that is missed by perturbative treatments.

Beyond the external barrier, the depth of the internal nuclear potential is 
another critical ingredient of the $\alpha$-decay cluster model. In the 
original Gamow picture, tunneling is dominated by the Coulomb barrier outside 
the nucleus ($r > R_0$), and the internal well is assigned a constant depth 
that plays only a minor role. The square-well radius $R_0$, however, is 
extremely sensitive: a mere 2\% variation in $R_0$ can change the calculated 
half-life $T_{1/2}$ by a factor of two. Recognizing this limitation, Buck, 
Merchant, and Perez proposed the BMP model in 1990~\cite{Buck90,Buck91}, 
which treats the decaying system as an ``$\alpha$-core'' configuration where 
the $\alpha$ particle occupies well-defined single-particle orbitals around the core.
In this framework, $R_0$ is determined self-consistently from the Bohr--Sommerfeld quantization condition for the internal 
$\alpha$-particle wave function at a given potential depth and node number, rather than being fixed by the empirical charge radius. The square-well depth $V_0$ thus becomes intrinsically linked to 
$R_0$, directly influencing the tunneling probability and half-life. Accurate predictions of $\alpha$-decay half-lives 
therefore require a fine-tuning of the nuclear potential depth for each 
individual nucleus---a procedure that naturally captures variations in the 
underlying shell structure. The central goal of the present work is therefore to develop an improved BMP model that (i) treats the centrifugal barrier analytically and non-perturbatively, and (ii) replaces the nucleus-by-nucleus fitted depth with a unified four-parameter formula applicable to both favored and unfavored transitions across the entire nuclear chart.

In our previous work~\cite{Tian24}, we refined the BMP model by replacing 
its constant nuclear potential depth with a structure-dependent form 
incorporating shell-correction energy, reducing the root-mean-square (rms) deviation between calculated and experimental half-lives by  32\% (from 0.278 to 0.188) for 178 even--even $\alpha$ emitters. 
That formulation, however, was restricted to even-even nuclei. Introducing separate parameterizations for odd-$A$ and odd--odd systems would substantially expand the parameter space, thereby undermining the goal of a unified description across the entire nuclear chart with a minimal set of parameters. To achieve such a unified framework, we incorporate odd--even 
staggering together with the orbital-angular-momentum dependence by 
classifying $\alpha$ transitions into favored ($L = 0$) and unfavored 
($L \neq 0$) categories. Favored decays serve as the reference set for the parameter calibration, upon which the systematic treatment of unfavored transitions is built.

\section{\label{sec:level2}Theoretical framework}

\subsection{\label{subsec:leve1} The Buck-Merchant-Perez (BMP) model and potential}
The $\alpha$-core interaction is the pivotal determinant of $\alpha$-decay 
half-lives within cluster models. It is conventionally partitioned into 
nuclear, Coulomb, and centrifugal components, among which the internal 
nuclear potential remains the least understood. Assuming a uniform nuclear density, we adopt the square-well form for the internal region. The effective $\alpha$-core potential thus reads
\begin{eqnarray}\label{Eq1}
V(r)=
\begin{cases}
-V_{0}=-V_{N}+\dfrac{C}{R}+\dfrac{\beta}{R^{2}}, & (r < R) \\[4pt]
\dfrac{C}{r}+\dfrac{\beta}{r^{2}}, & (r > R)
\end{cases}
\end{eqnarray}
where $C = Z_{\alpha}Z_{d}e^{2}$, with $Z_{\alpha}$ and $Z_{d}$ the proton 
numbers of the $\alpha$ particle and the daughter nucleus, and 
$\beta = \hbar^{2}L(L+1)/(2\mu)$ is the centrifugal coefficient. Here 
$\hbar \approx 197.327$~MeV$\cdot$fm/$c$, and 
$\mu = A_{\alpha}A_{d}\,{\rm u}/(A_{\alpha}+A_{d})$ is the reduced mass with 
${\rm u} = 931.494$~MeV/$c^{2}$. 

In contrast to Gamow’s original prescription, which assigns a constant depth $V(r) = -V_{0}$, the BMP model~\cite{Buck90} determines $V_{N}$ self-consistently through the Bohr--Sommerfeld quantization condition. This treatment incorporates the Coulomb and centrifugal contributions at the nuclear surface, thereby substantially improving the accuracy of half-life predictions. The square-well radius $R$ and depth $V_{N}$ are consequently coupled: for an $\alpha$ particle 
with orbital angular momentum $L$, global quantum number $G$, and decay 
energy $Q$, they satisfy
\begin{eqnarray}\label{eq:BS_quantization}
\int_{0}^{R}\sqrt{\frac{2\mu}{\hbar^{2}}\left[Q-V(r)\right]}dr=\frac{\pi}{2}(G-L+1).
\end{eqnarray}
Experimental $Q$ values are taken from the AME2020 mass table~\cite{AME2020}; for unmeasured cases, the 
$Q$ values are calculated using the WS3+ mass model~\cite{Wang11}.

Solving Eqs.~(\ref{Eq1}) and (\ref{eq:BS_quantization}) simultaneously gives
\begin{eqnarray}\label{Eq3}
R=\frac{C+\sqrt{C^{2}+4(Q+V_{N})\frac{\hbar^{2}}{2\mu}\{L(L+1)+\left[\frac{\pi}{2}(G-L+1)\right]^{2}}\}}{2\left(Q+V_{N}\right)}.
\end{eqnarray}
\emph{Crucially, $R$ depends on $L$ through Eq.~\eqref{eq:BS_quantization}}---a 
feature absent from the treatments of Zhang \emph{et al.} and 
Rojas-Gamboa \emph{et al.}, where $R$ is held fixed. Gamow captured the 
$L$-dependence perturbatively through $R_{\rm eff}$, but without enforcing a 
self-consistent quantization condition.

This radius provides a crucial input to the decay width, which in the 
semiclassical approximation~\cite{Gurv87} reads
\begin{eqnarray}\label{Eq4}
\Gamma=\frac{P_{\alpha}\hbar^{2}K}{2\mu R}{\rm exp}\left[-2\int_{R}^{R_{L}}k(r)dr\right],
\end{eqnarray}
so that the half-life is
\begin{eqnarray}\label{Eq8}
T_{1/2}=\frac{\hbar {\rm ln2}}{\Gamma}=\frac{2\mu R{\rm ln2}}{\hbar KP_{\alpha}}{\rm exp}\left[2\int_{R}^{R_{L}}k(r)dr\right].
\end{eqnarray}
Here $P_{\alpha}$ is the $\alpha$-particle preformation probability, set to 
unity; $K = \pi(G-L+1)/(2R)$ and 
$k(r) = \sqrt{(2\mu/\hbar^{2})\bigl(C/r + \beta/r^{2} - Q\bigr)}$ are the 
wave numbers in the internal and barrier regions, respectively; and 
$R_{L} = \bigl(C + \sqrt{C^{2}+4Q\beta}\bigr)/(2Q)$ is the outer turning 
point for unfavored decays (as shown in Fig.~\ref{fig1}).

\subsection{\label{subsec:leve2} Analytical solution of the centrifugal-action integral}
For \textbf{unfavored decays} in the presence of the effective (Coulomb 
plus centrifugal) barrier, the action integral admits a closed-form 
analytical expression when $Q > 0$ and $0 < R < R_{L}$:
\begin{align}
J_{L} &= \int_{R}^{R_{L}} k(r)\,dr
     = \sqrt{\frac{2\mu}{\hbar^{2}}}
       \int_{R}^{R_{L}} \sqrt{\frac{C}{r}+\frac{\beta}{r^{2}}-Q}\,dr 
       \notag \\[6pt]
&= \sqrt{\frac{2\mu}{\hbar^{2}}}
   \left\{ -S(R)
   + \sqrt{\beta}\,\ln\!\left(\frac{2\beta + CR + 2\sqrt{\beta}\,S(R)}{R\,\Delta}\right)
   + \frac{C}{2\sqrt{Q}}\!\left[\frac{\pi}{2} 
     + \arcsin\!\left(\frac{C-2QR}{\Delta}\right)\right] \right\},
\label{Eq9}
\end{align}
with $S(R) = \sqrt{-QR^{2}+CR+\beta}$ and $\Delta = \sqrt{C^{2}+4Q\beta}$. 
Remarkably, the result depends solely on the inner radius $R$ and is 
independent of the outer turning point $R_{L}$. A detailed derivation is 
given in Appendix~A.

For \textbf{favored decays} ($L = 0$), the centrifugal term vanishes 
($\beta = 0$) and the effective barrier reduces to the pure Coulomb form 
(black curve in Fig.~\ref{fig1}), with the outer turning point 
$R_{L} = R_{c} = C/Q$. The action integral 
then simplifies to
\begin{eqnarray}\label{Eq10}
J_{0} = \int_{R}^{R_{c}} k(r)\,dr
     = \sqrt{\frac{2\mu}{\hbar^{2}}}\,\frac{C}{\sqrt{Q}}
       \left[\arccos\!\left(\sqrt{\frac{R}{R_{c}}}\right)
       - \sqrt{\frac{R}{R_{c}} - \left(\frac{R}{R_{c}}\right)^{2}}\right],
\end{eqnarray}
recovering the \emph{standard Gamow penetration factor} for a pure Coulomb 
barrier~\cite{Gamow1928,Gamow1931,Tian24}, as expected.

Substituting the action integral in Eq.~(\ref{Eq9}) into Eq.~(\ref{Eq8}) and 
taking the common logarithm, we obtain an analytical decay formula of the 
Geiger--Nuttall type~\cite{Bayrak20},
\begin{equation}\label{Eq11}
\log_{10} T_{1/2} = a + b\, Q^{-1/2},
\end{equation}
where the coefficients $a$ and $b$ retain a residual dependence on $Q$ and 
$L$ through $S(R)$, $\Delta$, and $R$,
\begin{align}\label{Eq12}
a &= \log_{10}\!\left( 
     \frac{4\mu R^{2} \ln 2}{P_{\alpha}\, \pi \hbar (G - L + 1)} 
     \right) \notag \\[4pt]
  &\quad + 2 \sqrt{\frac{2\mu}{\hbar^{2}}} \left\{ 
     -S(R) + \sqrt{\beta}\, \ln\!\left( 
     \frac{2\beta + CR + 2\sqrt{\beta}\, S(R)}{R\Delta} \right) 
     \right\} \log_{10}(e), \\[6pt]
b &= C \sqrt{\frac{2\mu}{\hbar^{2}}} 
     \left[ \frac{\pi}{2} + \arcsin\!\left( 
     \frac{C - 2QR}{\Delta} \right) \right] \log_{10}(e).
\end{align}

\subsection{\label{subsec:leve3} Unified potential-depth formula incorporating the centrifugal term}

In this work we propose a compact formula for the nuclear potential depth 
$V_{N}$ that applies uniformly across different parity combinations, with 
parameters determined from individual $\alpha$-emitter properties within 
the BMP framework.
We first extract ``experimental'' potential depths $V_{N}^{\rm exp}$ from 
measured half-lives $T_{1/2}^{\rm exp}$ by numerically inverting 
Eqs.~(\ref{Eq3})--(\ref{Eq11}), assuming a constant preformation probability 
$P_{\alpha} = 1$ as in the BMP model. Our dataset comprises 534 half-lives 
of nuclei with $60 \le Z \le 118$, selected from the NUBASE2020 
database~\cite{Wangm21} with a relative uncertainty below 50\%. Detailed 
information for these nuclei---including potential depths and widths, 
shell-correction energies, $Q$ values, and both experimental and theoretical 
half-lives---is compiled in the Supplemental Material~\cite{Tian26}. These 
data characterize the $\alpha$-core interaction inside the parent nuclei 
and provide a reliable basis for predicting half-lives of hitherto 
unmeasured cases.

To construct a universal expression for $V_{N}$, we augment 
Eq.~(\ref{Eq1}) with a pairing term and an angular-momentum term:
\begin{equation}\label{Eq13}
V_{N}^{\rm fit} = c_{0} + \left(c_{1}A^{2} + c_{2}A 
                 + c_{3}E_{sh}\frac{AZ}{Q}\right)\!\Big/G
                 - 2\bigl[(-1)^{Z} + (-1)^{N}\bigr] - L(L+1),
\end{equation}
where $c_{0} = 229$~MeV, $c_{1} = 0.0255$~MeV, $c_{2} = -15.8277$~MeV, 
and $c_{3} = 0.0092$~MeV are obtained from a least-squares fit to 380 
favored $\alpha$-transitions ($L = 0$). 
Only ground-state-to-ground-state transitions enter the fit; 
nevertheless, the resulting systematics reproduce unfavored decays 
($L \ne 0$) remarkably well.

The pairing term $-2[(-1)^{Z} + (-1)^{N}]$ provides a unified description 
of all parity combinations and captures the well-known odd--even staggering 
in $\alpha$-decay, thus obviating separate parameterizations for even--even, 
odd-$A$, and odd--odd nuclei. Within an isotopic chain, the half-life of an 
even--even nucleus is typically shorter than the average of its odd-mass 
neighbors, whereas that of an odd--odd nucleus is longer. The pairing term 
translates this behavior into potential-depth shifts of approximately 
$-4$, $+4$, and $0$~MeV for even--even, odd--odd, and odd-$A$ 
nuclei, respectively, producing a shallower well for even--even systems 
and a deeper one for odd--odd systems.
The second additional term, $-L(L+1)$, accounts for the orbital angular 
momentum carried by the emitted $\alpha$ particle, yielding a shallower 
well for unfavored decays than for favored ones. Because the centrifugal 
barrier suppresses tunneling exponentially---far outweighing its weak 
linear influence on the initial well depth---increasing $L$ monotonically 
reduces the penetration probability and lengthens the half-life. 
Notably, when two independent coefficients $c_{4}$ and $c_{5}$ 
are introduced in front of these two terms, the fit returns 
$c_{4} \approx 2$~MeV and $c_{5} \approx 1$~MeV, supporting the present 
form.
The global quantum number $G$ is taken to be 22 for $N \le 126$ and 24 
for $N > 126$ in favored transitions, following Ref.~\cite{Buck90}, and 
$G = 24$ for all unfavored transitions.

\begin{figure}
\includegraphics[angle=-0,width= 1.02\textwidth]{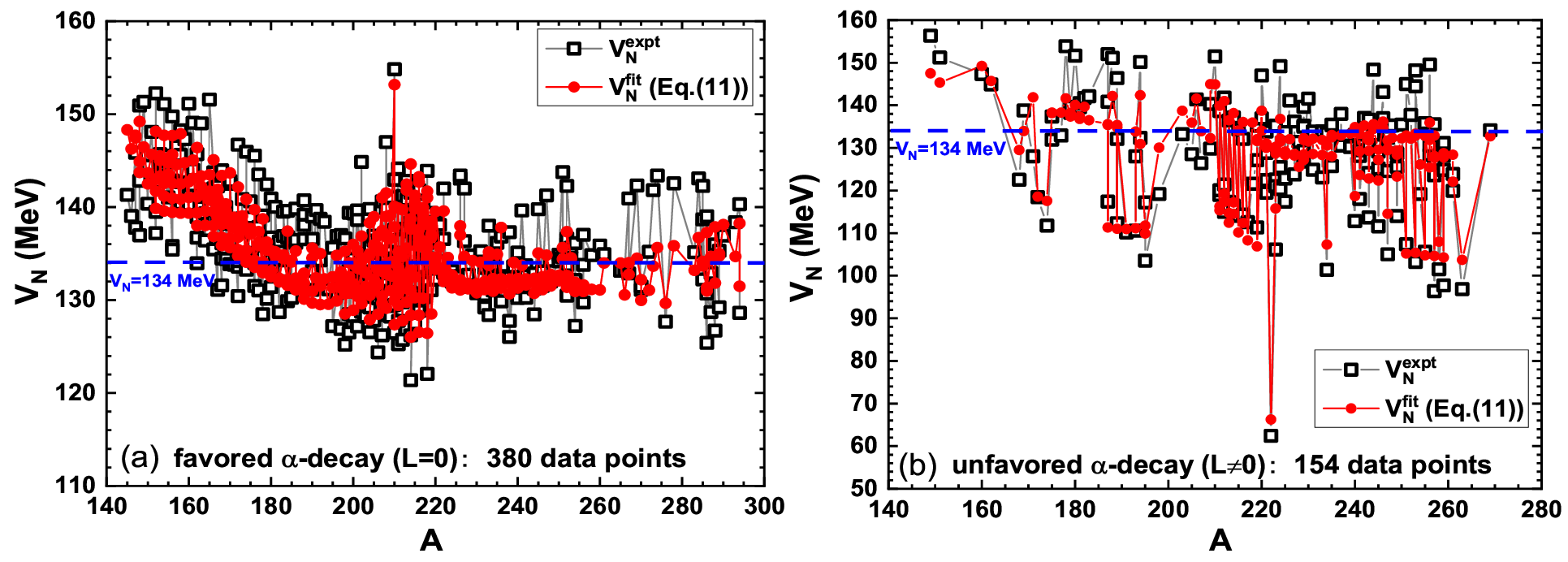}
 \caption{(Color online) Values of $V_{N}$ as a function of mass numbers $A$ of parent nuclei in different types of $\alpha$-decay (a) for favored (\(L = 0\)), analyzing 380 data points, and (b) unfavored (\(L \ne 0\)), analyzing 154 data points. The open squares denote the extracted potential depths $V_{N}^{\rm exp}$ from the measured half-lives. The solid circles denote the calculation results with Eq. (\ref{Eq13}). The average value of the open squares equals approximately to 134 MeV with dashed line marked.}\label{fig2}
\end{figure}
Figure~\ref{fig2} compares the experimental nuclear potential depths 
($V_{N}^{\rm exp}$, open squares) with the theoretical values from 
Eq.~\eqref{Eq13} ($V_{N}^{\rm fit}$, solid circles) for $\alpha$-emitting 
parent nuclei over a broad mass range. Panel~(a) shows the 380 favored 
transitions ($L=0$): $V_{N}^{\rm exp}$ decreases smoothly with $A$ and 
exhibits only moderate scatter, indicating good agreement between theory 
and experiment. Panel~(b) shows the 154 unfavored transitions ($L\ne 0$); 
the same overall trend persists, though the scatter is larger---especially 
in the region $A \approx 180$--$250$---reflecting the enhanced sensitivity 
of unfavored decays to nuclear-structure effects. 
Notably, the smallest value of $V_{N}$ occurs at $A=222$, 
$Z=93$, where the emitted $\alpha$ particle carries $L=8$---the largest 
angular momentum among the 154 unfavored transitions considered. The 
strong centrifugal hindrance in this case yields a markedly shallower 
effective well within the present description.

The dashed line in Fig.~\ref{fig2} marks the mean value 
$\overline{V_{N}^{\rm exp}} = 134$~MeV, in accord with the constant depth 
adopted in the original BMP model~\cite{Buck90}. This agreement is 
striking: it holds for both favored and unfavored decays and spans light 
to heavy nuclei ($A \approx$ 140 -- 300). At the same time, the visible 
deviations from the mean demonstrate that a constant depth cannot capture 
the full nucleus-dependent structural information.
A prominent feature is the sharp spike in $V_{N}^{\rm exp}$ at the 
$N=126$ shell closure, e.g., $^{210}$Pb with $V_{N}^{\rm exp} = 154.741$~MeV, which 
highlights pronounced shell effects. Since it is impractical to account 
for every individual nuclear property, we instead incorporate the shell 
correction energy $E_{\rm sh}$ of the parent nucleus. This quantity can 
be taken from mass tables or evaluated as the difference between the 
experimental and macroscopic binding energies:
\begin{equation}\label{Eq14}
E_{sh}(A,Z)=B_{\rm exp}(A,Z)-B_{\rm m}(A,Z).
\end{equation}
The smooth liquid-drop energy $B_{\rm m}$ of a spherical nucleus is given 
by a modified Bethe--Weizs\"acker mass formula~\cite{Wangn14},
\begin{equation}\label{Eq15}
B_{m}(A,Z)=a_{v}A-a_{s}A^{2/3}-a_{c}\frac{Z^{2}}{A^{1/3}}(1-0.76Z^{-2/3})-a_{sym}I^{2}Af_{s},
\end{equation}
with the symmetry-energy coefficient $a_{sym}=c_{sym}(1-\frac{\kappa}{A^{1/3}}+\xi\frac{2-|I|}{2+|I|A})$
and the surface-diffuseness correction factor 
$f_{s} = 1 + \kappa_{s}\,\epsilon\,A^{1/3}$. Here 
$\epsilon = (I - I_{0})^{2} - I^{4}$ encodes the deviation from the 
constant surface diffuseness of the Woods--Saxon potential, and 
$I_{0} = 0.4A/(A+200)$ is the isospin asymmetry along the 
$\beta$-stability line as given by Green's formula. The parameters are 
$a_{v} = 15.5181$~MeV, $a_{s} = 17.4090$~MeV, $a_{c} = 0.7092$~MeV, 
$c_{\rm sym} = 30.1594$~MeV, $\kappa = 1.5189$, $\xi = 1.2230$, and 
$\kappa_{s} = 0.1536$.

These results demonstrate that when $V_{N}$ is expressed as a function of 
fundamental quantities---the mass number $A$, charge number $Z$, shell 
correction energy $E_{\rm sh}$, and $Q$ value---it effectively captures 
the structural effects of the parent nuclei.
\begin{figure}
\includegraphics[angle=-0,width= 0.75\textwidth]{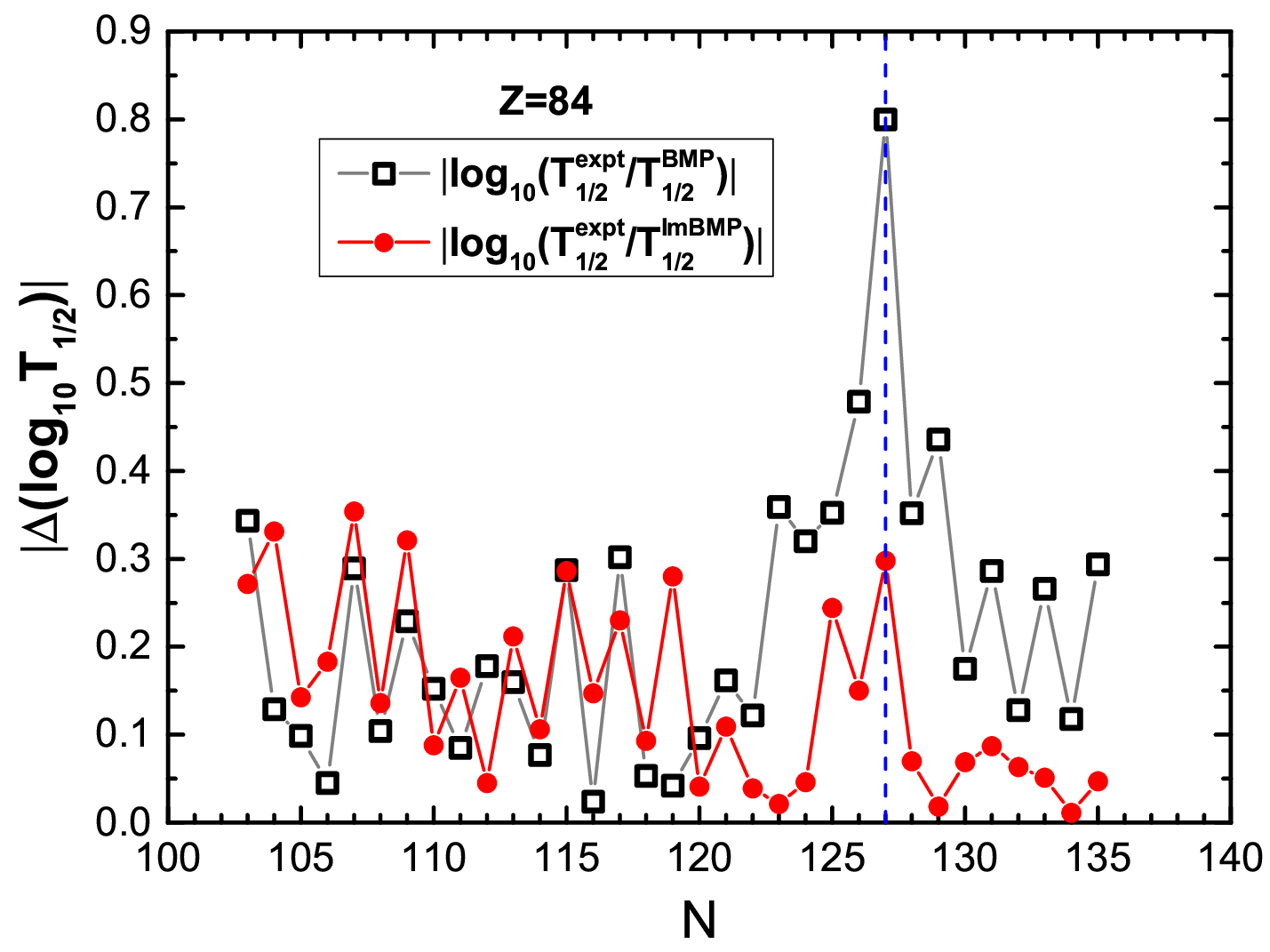}
 \caption{(Color online) Absolute value of logarithms of the ratios between theoretical $\alpha$-decay half-lives and experimental ones for Po isotopic chain with the BMP model (open squares) and with Eq. (\ref{Eq13}) (solid circles) for $V_{N}$ in the ImBMP model.}\label{fig3}
\end{figure}

\section{\label{sec:level3}The results and discussions}
Figure~\ref{fig3} shows the absolute logarithmic deviation 
$|\Delta(\log_{10}T_{1/2})|$ along the Po isotopic chain. The open squares 
represent the results of the original BMP model with $V_{N}=134$~MeV, 
whereas the solid circles correspond to the ImBMP results calculated from 
Eq.~(\ref{Eq13}). Here, $\Delta ({\rm log}_{10}T_{1/2})$ = ${\rm log}_{10}(T_{1/2}^{\rm expt}/T_{1/2}^{\rm cal})$
measures the deviation between calculated and experimental $\alpha$-decay 
half-lives on a logarithmic scale. As seen in Fig.~\ref{fig3}, the BMP 
model exhibits a pronounced deviation around the neutron magic number 
$N=126$ in the Po isotopic chain. By contrast, the values obtained from 
Eq.~(\ref{Eq13}) are in much better agreement with the experimental data 
near $N=126$. A similar trend is also found in other isotopic chains 
crossing the neutron magic number 126. This systematic behavior appears to 
be closely related to shell structure, indicating that incorporating shell 
corrections into the potential depth is important for a more accurate 
description of $\alpha$-decay half-lives.
\begin{figure}
\includegraphics[angle=-0,width= 1.01\textwidth]{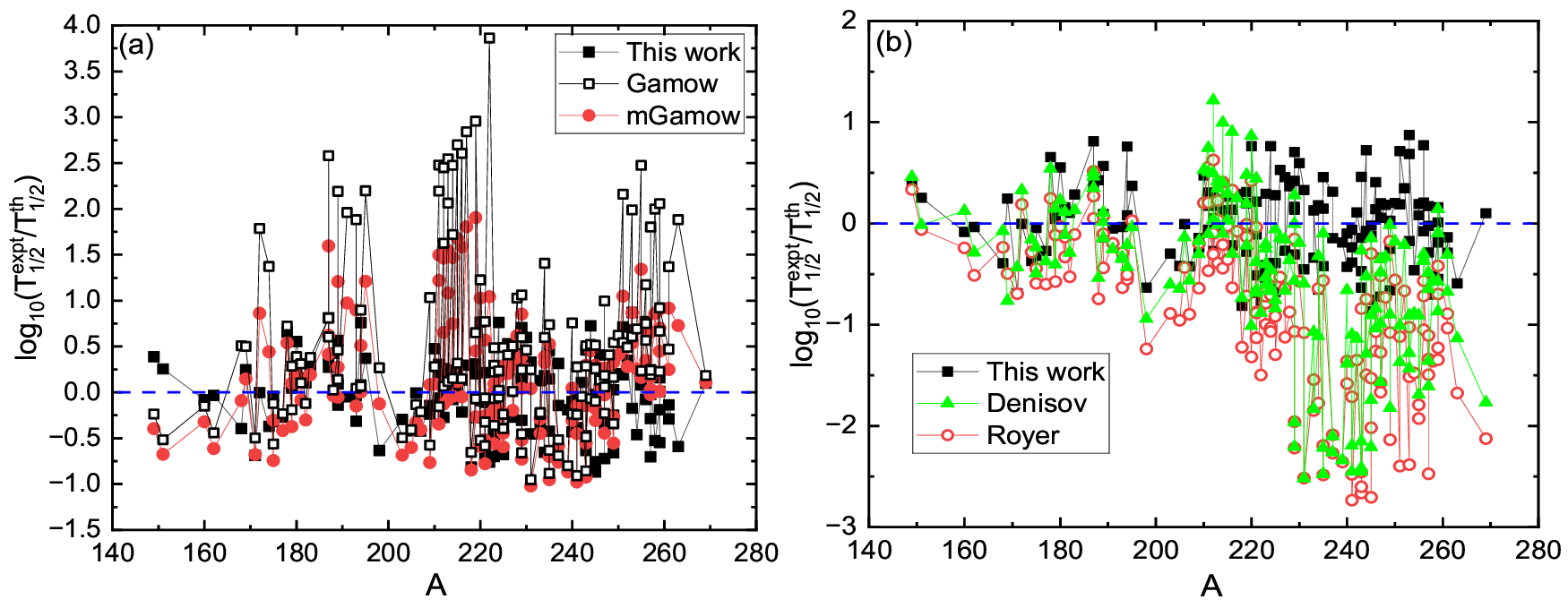}
 \caption{(Color online) Comparison of the logarithmic deviations between calculated and experimental half-lives for 154 unfavored $\alpha$ emitters as functions of the parent mass number $A$. Panel (a) compares the present ImBMP results with those obtained from the original Gamow formula and the modified Gamow prescription. Panel (b) compares the present results with those from the Denisov~\cite{Denisov09} and Royer~\cite{Royer10} formulas.}\label{fig4}
\end{figure}

To further clarify the role of the centrifugal potential in unfavored $\alpha$ decay, we compare the present ImBMP model with several representative schemes for 154 unfavored $\alpha$ emitters. Fig.~\ref{fig4}(a) compares the present results with those from the original Gamow formula and a modified Gamow prescription. The Gamow formula reads:
\begin{equation}\label{Eq16}
\log_{10}T_{1/2} = -21.8285 + 2 \log_{10}R_{0} + 1.7185\frac{Z_{d}}{\sqrt{Q}} -1.2895\sqrt{\frac{A_{d}Z_{d}R_{0}}{A}}
\left[1-0.001L(L+1)\frac{A}{A_{d}}\right].
\end{equation}
When applied directly with a fixed radius parameter $R_{0}$, this formula yields sizable deviations from experimental half-lives, indicating that a constant radius cannot adequately describe the centrifugal hindrance. Replacing $R_{0}$ with an effective radius $R_{\mathrm{eff}}=R_{0}[1-0.002L(L+1)]$, where $R_{0}$ is determined from the Bohr--Sommerfeld quantization condition for $L=0$ in the BMP framework, significantly reduces the deviations [red solid circles in Fig.~\ref{fig4}(a)]. This improvement shows that the centrifugal effect modifies not only the barrier height but also the effective radial scale of the tunneling problem. Nevertheless, the modified Gamow prescription remains approximate, as the centrifugal contribution is introduced only through an effective-radius substitution. In contrast, the ImBMP model incorporates the centrifugal potential directly into the $\alpha$-core potential, determines the radius self-consistently, and evaluates the WKB penetration integral analytically for the Coulomb-plus-centrifugal barrier, yielding a more consistent description of both barrier height and width.

For a broader benchmark, we also compare our results with the widely used Denisov~\cite{Denisov09} and Royer~\cite{Royer10} formulas:
\begin{equation}\label{Eq17}
\log_{10}T_{1/2} = a + b \frac{A^{1/6} Z^{1/2}}{\mu’} + \frac{cZ}{\sqrt{Q}}+
\frac{d \sqrt{L (L + 1)}}{Q A^{-1/6}} + e \left[(-1)^L - 1\right],
\end{equation}
and
\begin{equation}\label{Eq18}
\log_{10} T_{1/2} = a + b A^{1/6} Z^{1/2} + \frac{cZ}{\sqrt{Q}}+
\frac{dAZN [L (L + 1)]^{1/4}}{Q} + eA[1 -(-1)^L],
\end{equation}
where $A$, $Z$, and $N$ denote the mass, charge, and neutron numbers of the parent nucleus, $\mu' = [A/(A-4)]^{1/6}$ in Eq.~(\ref{Eq17}), and the parameters $a$, $b$, $c$, $d$, and $e$ are fitted to experimental data~\cite{Denisov09,Royer10}. In these formulas, the angular-momentum dependence is incorporated through phenomenological hindrance terms involving $L(L+1)$ or its fractional power, treating the centrifugal effect mainly as a fitted correction rather than through explicit modification of the turning points and WKB action.

In the ImBMP model, the $L$ dependence enters directly through the $\alpha$-core potential and the barrier-penetration integral. The centrifugal potential is included explicitly in the effective barrier, the inner radius is determined self-consistently from the Bohr–Sommerfeld quantization condition, and the potential depth $V_{N}$ incorporates shell-correction, odd–even, and angular-momentum terms via Eq.~(\ref{Eq13}). As shown in Fig.~\ref{fig4}(b), the deviations obtained in the present calculation are more tightly centered around zero than those from the Denisov and Royer formulas for the 154 unfavored $\alpha$ emitters. This improvement indicates that the centrifugal hindrance is not simply an empirical additive correction, but is intimately connected with the modification of the barrier geometry and the nuclear-structure dependence of the internal $\alpha$-core potential.

\begin{figure}
\includegraphics[angle=-0,width= 0.95\textwidth]{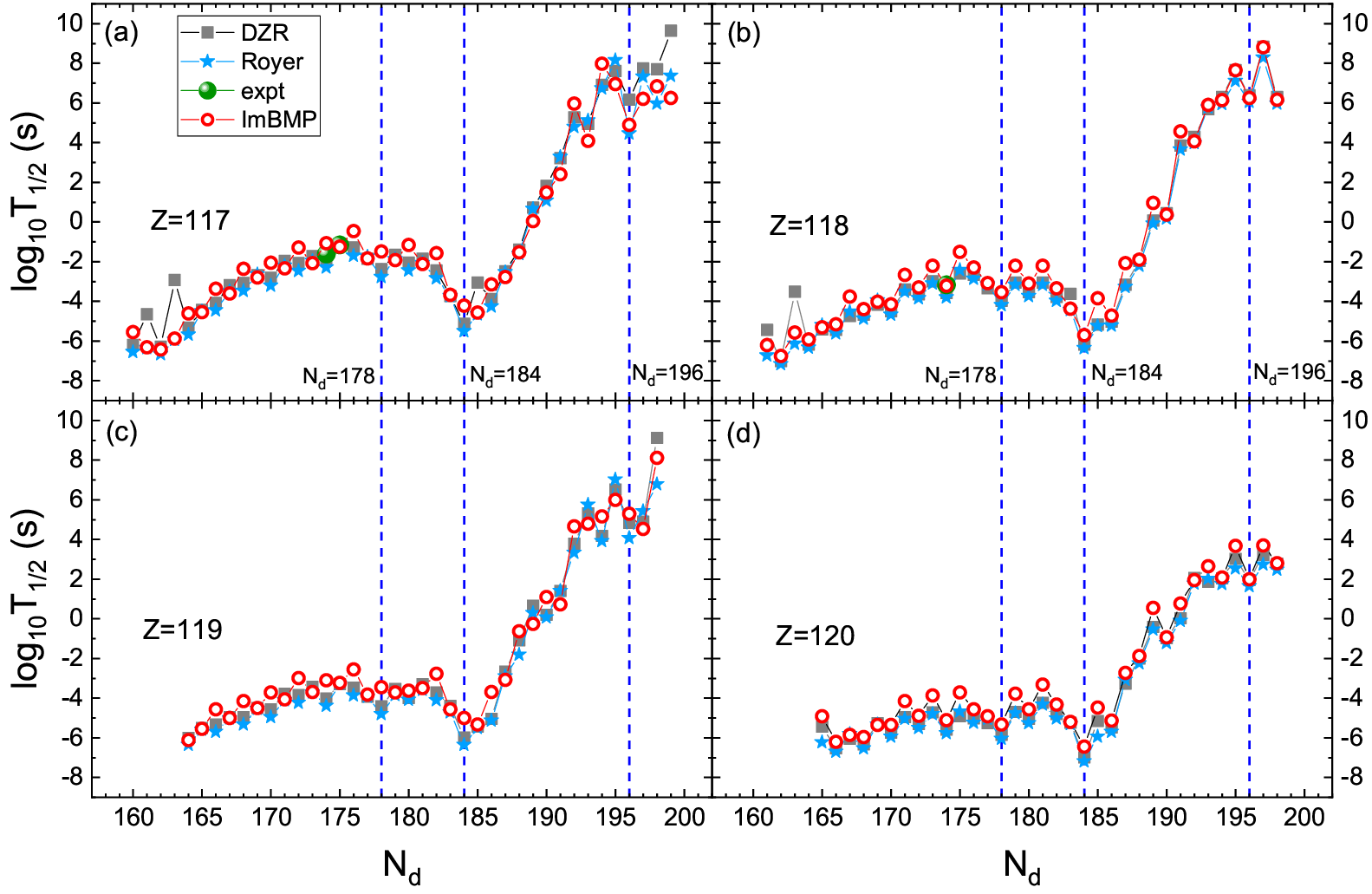}
 \caption{(Color online) $\log_{10}T_{1/2}$ values for the 
 $Z=117$--$120$ isotopic chains, shown in panels (a)--(d), as functions of 
 the daughter neutron number $N_{d}$. Open circles denote the predictions 
 of the ImBMP model. Black squares and blue stars correspond to the 
 Deng-Zhang-Royer (DZR) formula proposed by Deng \emph{et al.}~\cite{Deng20} 
 and the original Royer formula~\cite{Royer10}, respectively. Solid green circles indicate 
 the experimental data for $^{293,294}\mathrm{Ts}$ and $^{294}\mathrm{Og}$.}\label{fig5}
\end{figure}

Figure~\ref{fig5} presents the evolution of $\log_{10}T_{1/2}$ 
for the four isotopic chains with $Z=117$--$120$ as functions of the 
daughter neutron number $N_{d}$. For comparison, predictions from the 
original Royer formula~\cite{Royer10} and the Deng-Zhang-Royer (DZR) 
formula~\cite{Deng20} are also shown. To ensure a consistent comparison, the unmeasured $\alpha$-decay energies used here are calculated from the same WS3+ mass model~\cite{Wang11}, and the orbital angular momenta $L$ of the $\alpha$ particle are taken from Denisov~\cite{Denisov25}. The ImBMP predictions proposed in 
this work are in excellent agreement with those of the DZR and Royer 
formulas. Moreover, among the models considered, the ImBMP results show 
the best overall agreement with the available experimental data (solid 
green circles), confirming the reliability and accuracy of the present 
approach for describing $\alpha$ decay in heavy and superheavy nuclei.
Pronounced shell effects are also observed at $N_{d}=178$, 184, and 196 
along all four isotopic chains, providing useful signatures for possible shell or subshell in this superheavy region. Although the absolute values of the calculated $\alpha$-decay half-lives depend on the theoretical framework adopted, the consistent indication of shell-closure locations by different models underscores the robustness of the underlying nuclear-structure effects.

\section{\label{sec:level4}Summary}
In summary, this work has presented an improved 
Buck--Merchant--Perez (BMP) model for $\alpha$-decay half-lives, 
built upon two complementary developments: a closed-form, 
non-perturbative WKB penetration formula that incorporates the 
centrifugal potential exactly, and a unified four-parameter formula 
for the nuclear potential depth $V_{N}$ that accounts for shell 
corrections, odd--even pairing effects, and orbital-angular-momentum 
dependence. With only four adjustable parameters fitted to 380 
favored ground-state-to-ground-state transitions, the same formula 
reproduces 154 unfavored transitions without any further adjustment, 
thereby removing the need for separate parametrizations of odd-$A$ 
and odd--odd systems. For the full set of 534 nuclei spanning 
$60 \le Z \le 118$, the calculated $\log_{10}T_{1/2}$ values yield 
a root-mean-square deviation of  0.267, a 57\% 
improvement over the original constant-depth BMP model 
(0.615). The framework is further applied to 
predict the $\alpha$-decay half-lives of superheavy nuclei with 
$Z = 117$--$120$, and the results are in good agreement with those 
of other established approaches, providing quantitative benchmarks 
for future experimental investigations of the superheavy region.

\begin{center}
\textbf{ACKNOWLEDGMENTS}
\end{center}
This work was supported by the Guangxi Science and Technology Program (No. 2023GXNSFDA026005 and No. 2023GXNSFBA026008), the National Natural Science Foundation of China (No. 12465019 and No. 12465021), and the Central Government Guides Local Scientific and Technological Development Fund Projects (No. Guike ZY22096024).

\appendix

\section{Analytical Integral with the Centrifugal Potential}
\label{app:integral_derivation}
We consider the barrier penetration problem in nuclear reactions. The inner region ($r \leq R$) is modeled as a square--well, while the outer region ($r > R$) consists of the Coulomb potential superimposed with the centrifugal potential. Since the penetration integral involves only the classically forbidden region $R < r < R_L$, we employ the outer-region potential, which can be written as
\begin{equation}
    V(r) = \frac{Z_\alpha Z_d e^2}{r} + \frac{\hbar^2 L(L+1)}{2\mu r^2} \equiv \frac{C}{r} + \frac{\beta}{r^2},
    \label{eq:potential}
\end{equation}
where $C = Z_\alpha Z_d e^2$ and $\beta = \hbar^2 L(L+1)/(2\mu)$.

In the WKB approximation, the barrier penetration probability is given by
\begin{equation}
    P_L = \exp(-2K_L),
    \label{eq:penetration_prob}
\end{equation}
with
\begin{equation}
    K_L = \frac{\sqrt{2\mu}}{\hbar}\int_R^{R_L}\sqrt{V(r)-Q}\,dr 
    = \frac{\sqrt{2\mu}}{\hbar}\int_R^{R_L}\frac{S(r)}{r}\,dr\equiv \frac{\sqrt{2\mu}}{\hbar} J,
    \label{eq:WKB_action}
\end{equation}
where $S(r)=\sqrt{-Qr^2+Cr+\beta}$.

The central task is to evaluate the integral
\begin{equation}
    J = \int_{R}^{R_L}\frac{S(r)}{r}\,\mathrm{d}r,
    \label{eq:integral_J}
\end{equation}
we first seek the indefinite integral. Through using the standard integration formula and referring to the "Table of Integrals" (Gradshteyn and Ryzhik)~\cite{Gradshteyn2007}, we propose the antiderivative
\begin{equation}
    \mathcal{F}(r) = S(r) - \sqrt{\beta}\ln\left(\frac{2\beta + Cr + 2\sqrt{\beta}\,S(r)}{r}\right) - \frac{C}{2\sqrt{Q}}\arcsin\left(\frac{C-2Qr}{\Delta}\right),
    \label{eq:antiderivative}
\end{equation}
where $\Delta=\sqrt{C^2+4Q\beta}$ and the integration constant has been omitted. We now evaluate the definite integral
$J = \mathcal{F}(R_L) - \mathcal{F}(R)$.

\subsection{Value at the Upper Limit $r = R_L$}

The outer turning point $R_L$ is determined by $V(R_L) = Q$:
\begin{equation}
    QR_L^2 - CR_L - \beta = 0.
    \label{eq:turning_point}
\end{equation}
Taking the positive root yields
\begin{equation}
    R_L = \frac{C + \sqrt{C^2 + 4Q\beta}}{2Q} = \frac{C + \Delta}{2Q}.
    \label{eq:R_L}
\end{equation}

From Eq.~\eqref{eq:turning_point}, we immediately have $S(R_L) = 0$. Furthermore,
\begin{equation}
    \frac{C-2QR_L}{\Delta} = \frac{C-(C+\Delta)}{\Delta} = -1 
    \quad\Longrightarrow\quad 
    \arcsin\left(\frac{C-2QR_L}{\Delta}\right) = -\frac{\pi}{2}.
    \label{eq:arcsin_at_RL}
\end{equation}

From Eq.~\eqref{eq:R_L}, we also derive the key relation
\begin{equation}
    \Delta = 2QR_L - C = \frac{2\beta + CR_L}{R_L}.
    \label{eq:Delta_relation}
\end{equation}
Hence, the antiderivative at the upper limit evaluates to
\begin{align}
    \mathcal{F}(R_L) 
    &= 0 - \sqrt{\beta}\ln\Delta - \frac{C}{2\sqrt{Q}}\left(-\frac{\pi}{2}\right) \notag\\
    &= -\sqrt{\beta}\ln\Delta + \frac{C\pi}{4\sqrt{Q}}.
    \label{eq:F_RL}
\end{align}

\subsection{Value at the Lower Limit $r = R$}

At the lower limit, we have
\begin{equation}
    \mathcal{F}(R) = S(R) - \sqrt{\beta}\ln\left(\frac{2\beta + CR + 2\sqrt{\beta}\,S(R)}{R}\right) - \frac{C}{2\sqrt{Q}}\arcsin\left(\frac{C-2QR}{\Delta}\right).
    \label{eq:F_R}
\end{equation}

\subsection{Final Result}

Subtracting Eq.~\eqref{eq:F_R} from Eq.~\eqref{eq:F_RL} and rearranging, we obtain
\begin{align}
    J &= \mathcal{F}(R_L) - \mathcal{F}(R) \notag\\
    &= -S(R) + \sqrt{\beta}\ln\left(\frac{2\beta + CR + 2\sqrt{\beta}\,S(R)}{R\Delta}\right) + \frac{C}{2\sqrt{Q}}\left[\frac{\pi}{2} + \arcsin\left(\frac{C-2QR}{\Delta}\right)\right] 
    \label{eq:J_final_explicit}
\end{align}
This is Eq.~\eqref{Eq9} of the main text.

Substituting into Eq.~\eqref{eq:penetration_prob}, the exact penetration 
probability is:
\begin{align}
P_l = \left[\frac{R\Delta}{2\beta + CR + 2\sqrt{\beta}S(R)}\right]^{2\sqrt{L(L+1)}}
\exp\left\{\frac{2\sqrt{2\mu}}{\hbar}S(R) - \frac{C}{\hbar}\sqrt{\frac{2\mu}{Q}}\left[\frac{\pi}{2} + \arcsin\left(\frac{C-2QR}{\Delta}\right)\right]\right\}.
\label{eq:app_exact_PL}
\end{align}
\emph{Equation~\eqref{eq:app_exact_PL} is exact within the WKB 
approximation}---no additional approximation has been introduced beyond the 
standard WKB treatment. It is therefore equivalent to direct numerical 
integration for the same potential.

This appendix has presented a rigorous derivation of the barrier penetration integral in two stages: establishing the physical model and reducing the integral to a standard form (Sec.~\ref{subsec:leve2}) and evaluating the definite integral to obtain an analytical expression. It is noteworthy that this result depends only on the square-well radius $R$ and is independent of the outer turning point $R_L$---a consequence of the specific boundary behavior of the integrand at $R_L$.


\end{document}